# Object recognition in primates: What can early visual areas contribute?

**Christian Quaia and Richard J Krauzlis**

Laboratory of Sensorimotor Research, National Eye Institute, NIH, Bethesda, MD, USA

**If neuroscientists were asked which brain area is responsible for object recognition in primates, most would probably answer infero-temporal (IT) cortex. While IT is likely responsible for fine discriminations, and it is accordingly dominated by foveal visual inputs, there is more to object recognition than fine discrimination. Importantly, foveation of an object of interest usually requires recognizing, with reasonable confidence, its presence in the periphery. Arguably, IT plays a secondary role in such peripheral recognition, and other visual areas might instead be more critical. To investigate how signals carried by early visual processing areas (such as LGN and V1) could be used for object recognition in the periphery, we focused here on the task of distinguishing faces from non-faces. We tested how sensitive various models were to nuisance parameters, such as changes in scale and orientation of the image, and the type of image background. We found that a model of V1 simple or complex cells could provide quite reliable information, resulting in performance better than 80% in realistic scenarios. An LGN model performed considerably worse. Because peripheral recognition is both crucial to enable fine recognition (by bringing an object of interest on the fovea), and probably sufficient to account for a considerable fraction of our daily recognition-guided behavior, we think that the current focus on area IT and foveal processing is too narrow. We propose that rather than a hierarchical system with IT-like properties as its primary aim, object recognition should be seen as a parallel process, with high-accuracy foveal modules operating in parallel with lower-accuracy and faster modules that can operate across the visual field.**

## Introduction

It is commonly accepted that in primates visual recognition of an object is strongly dependent on the activity of neurons in the inferior temporal (IT) cortex (Tanaka, 1996; Logothetis and Sheinberg, 1996), an area at the end of the ventral visual stream (Ungerleider and Mishkin, 1982; Goodale and Milner, 1992). This hypothesis was first prompted by the discovery in monkey IT of neurons that respond vigorously to complex object shapes and faces (Gross et al., 1972; Desimone et al., 1984; Gross, 2008). Comparisons between the activity of artificial neural networks (ANNs) trained on object classification and the activity of large populations of IT neurons have lent further support to this hypothesis (Yamins and DiCarlo, 2016; Rajalingham et al., 2018; Yamins et al., 2014;

Khaligh-Razavi and Kriegeskorte, 2014; Kriegeskorte, 2015). Additional support came from the observation that large bilateral lesions of IT in monkeys result in considerable deficits in performing fine discriminations (Setogawa et al., 2021; Matsumoto et al., 2022; Eldridge et al., 2018). Similarly, causal manipulations of monkey IT activity while the animal is performing difficult discriminations impairs performance, further indicating a causal involvement of IT (Afraz et al., 2006; Moeller et al., 2017; Afraz et al., 2015).

Visual object recognition is however not limited to such fine discriminations. It is also important for quickly identifying and locating a potential prey or predator, thus providing information critical for survival. These needs are shared by animals that do not have the high visual acuity of diurnal primates, or anything resembling the ventral visual pathway (Kirk and Kay, 2004; Vinken et al., 2016;

Kaas, 2020; Leopold et al., 2020), and can usually be satisfied by a coarser type of recognition (enabling, for example, to tell apart a mouse from a cat). Just as importantly, fine discriminations are supported almost exclusively by foveal vision. Accordingly, IT has a strong foveal bias (Arcaro et al., 2017; De Beeck and Vogels, 2000; Op de Beeck et al., 2019), meaning that its neurons respond preferentially to images placed on or near the fovea. Initial reports seemed to indicate that IT had large receptive fields covering most of the contralateral visual field, but that was confounded by the use of large stimuli. When small stimuli are used, moving the stimulus by as little as 2 degrees away from the fovea results in a dramatic reduction in activity, even for preferred objects (DiCarlo and Maunsell, 2003; Hung et al., 2005). Taken together these observations suggest that IT in primates might be specialized for fine, mostly foveal, discriminations. This would readily explain the rather mild deficits in coarse object classification (e.g., dog versus cat) observed following bilateral ablations of IT in monkeys (Matsumoto et al., 2016).

Setting aside the needs of afoveated animals, we now end up with a chicken-and-egg problem: If IT is crucial for fine discriminations, and it relies predominantly on foveal inputs, how can an object of interest be foveated in the first place? Since our visual field covers approximately 180 deg x 120 deg, and the fovea is approximately 2 deg across, it would take 5400 refixations - 18 minutes at a rate of 5 saccades/s - to scan the entire visual field. Since this is not our daily experience (Potter, 1975; Drewes et al., 2011; Najemnik and Geisler, 2008), there must exist a system that is capable of locating objects in the environment and prioritizing those of likely behavioral interest. Importantly, this system must provide accurate information about the position of the target object, so that foveation can be achieved in a single saccade. This would require specificity for retinotopic location, thus eschewing the position invariance sought by ANNs.

The need for such a system - often involving a saliency or priority map - has been long recognized (Treisman, 1986; Serences and Yantis, 2006;

Fecteau and Munoz, 2006). The saliency map is usually envisioned as a topographically organized neural map of visual space, in which activity at any given location represents the priority to be given to that location for covert or overt shifts of attention (Itti et al., 1998). Originally, bottom-up saliency (e.g., local contrast, orientation, etc.) was considered the main determinant of this map (Itti and Koch, 2000), but it has become clear that the identification of objects (Einhäuser et al., 2008), and even more complex cognitive processes (Schütz et al., 2011), play a major role. Its neural substrate has been hypothesized to lie in a network of interconnected areas involving the superior colliculus (SC) in the midbrain (McPeek and Keller, 2002; White et al., 2017), the pulvinar nuclei in thalamus (Robinson and Petersen, 1992), the frontal eye fields (FEF) in prefrontal cortex (Thompson and Bichot, 2005), and the lateral intraparietal (LIP) area in parietal cortex (Gottlieb et al., 1998). The SC is likely to play a pivotal role, as it has a finer topographical organization than FEF and LIP, and it is more directly involved in controlling gaze and attention shifts (Krauzlis, 2005; Krauzlis et al., 2013). It also receives retinotopic inputs from most of the visual cortex, including areas V1, V2, V3, V4 and IT (Fries, 1984; Cusick, 1988; May, 2006). As noted, IT has a strong foveal bias, and presumably contributes less to the processing of peripheral objects and events than areas V1-V4. Topographically organized inputs from these areas to the salience network might then form the basis of a spatially localized system for peripheral (and coarse) object detection.

We thus sought to evaluate the ability of the first stages of cortical visual processing, simple and complex cells in area V1, to convey information about object category. We focused on information that can be extracted with a simple linear classifier (i.e., a weighted sum followed by a threshold operation, a good approximation of the operation performed by a neuron downstream of V1, (Rosenblatt, 1962)). We restricted our model to using only relatively low spatial frequencies, thus approximating processes in peripheral vision (Robson and Graham, 1981) or in infancy (Sokol, 1978). Finally, we focused on the classification task of identifying a face among non-face images. This is a coarse recognition task of strong behavioral significance for primates, and at which they are highly proficient (Diamond and Carey, 1986; Carey, 1992; Pascalis and Bachevalier, 1998; Pascalis et al., 1999; Rosenfeld and Van Hoesen, 1979; Tanaka and Gauthier, 1997), even when very young (Nelson and Ludemann, 1989; Nelson, 2001; Pascalis and de Schonen, 1994), and even in the (near) periphery (Mäkelä et al., 2001; McKone, 2004). To investigate the plausibility of V1 being directly involved in the detection of faces in the periphery, we assessed how variations in scale, orientation, and background affected the ability of a V1-like model to provide information useful to classification.

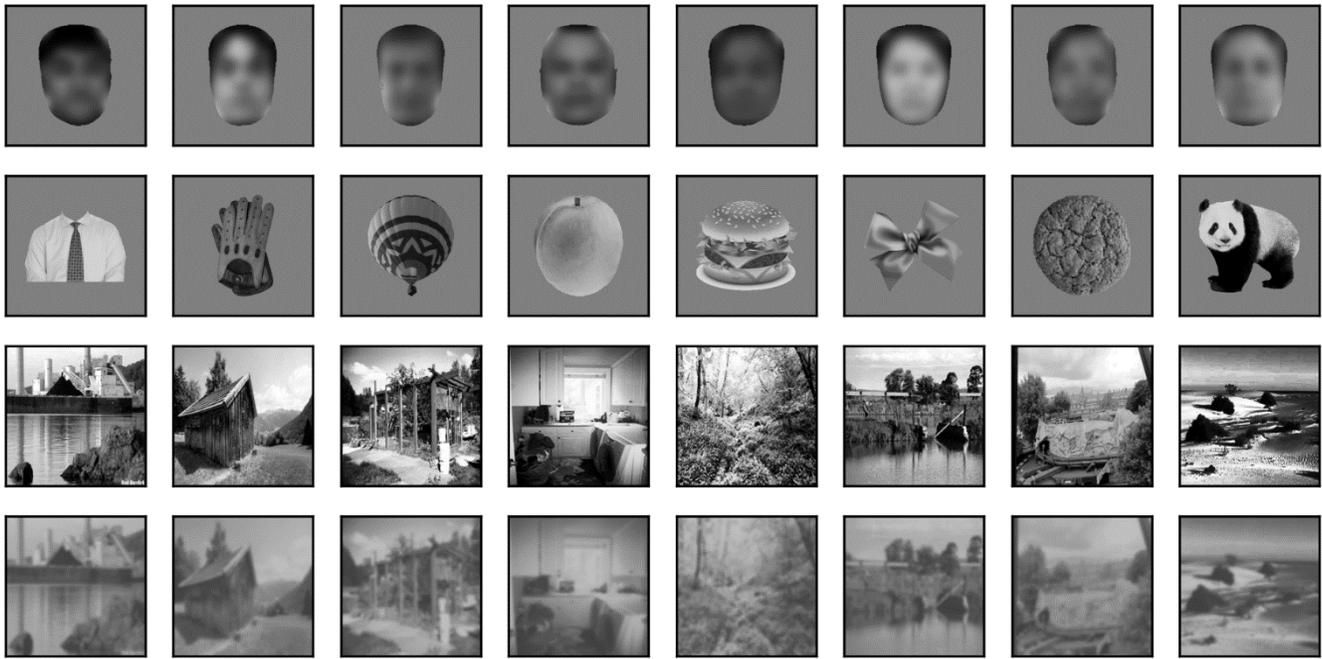

**Figure 1.** Random samples of images from sets. Top row: 8 images from our set of 1756 face images. Variations in sex, age, gender, and race, as well as contrast and mean luminance, seen above are representative of the entire set. Here, but not in the simulations, the images have been blurred to prevent identification of the subject. Second row: 8 images from our set of 1502 non-face images. Variations in shape, contrast and mean luminance seen above are representative of the entire set. Third row: 8 images from our set of 1525 high contrast background scenes. Variations in content, contrast and mean luminance seen above are representative of the entire set. Bottom row: Same 8 background images, but blurred and with their contrast reduced.

## Methods

### Image sets

For our face vs non-face image classification task, we assembled three sets of images: A set of 1,756 frontal images of human faces, a set of 1,502 images of non-face stimuli (which includes a variety of human-made objects, animals, fruits, and vegetables), and a set of 1,525 images of urban and natural landscapes.

We derived the face dataset from a larger (2,000 images) set of frontal images of faces that was kindly shared with us by Prof. Doris Tsao. The set contained images of subjects from both sexes and from all races, under varying levels of illumination. We modified this set in a few ways. First, we dropped 244 images that either had a lower resolution compared to the others or in which a large fraction of the face was occluded by hair or face coverings. Next, we resized each image from the original 256 x 360 (width x height) pixels (px) to our desired square size (256 x 256 px). This was done by preserving the aspect ratio and adding white space on the left and right sides. The image was then saved as a grayscale image with a transparency (alpha) channel (an RGBA image in which the red, green, and blue channels all have the same value). Because most images also contained hair, neck and shoulder, we used the dlib library (King, 2009) and the MTCNN face segmentation neural network (Zhang et al., 2016) to isolate the face region, removing completely neck and shoulders and leaving only minimum hair (while retaining an oval shape). Anything outside the face region was then set to transparent. Finally, the face region was rescaled (preserving its aspect ratio) to make its largest side equal to 246 px, and it was centered on the 256 x 256 px canvas. 8 samples from this set, centered on a 340 x 340 px mid gray background, are shown in Figure 1 (top row). Note that to make the subjects not identifiable we have blurred their images in Figures 1 and 4. The images presented to the models were however not blurred. Note that to make the subjects not identifiable we have blurred their images in Figures 1 and 4. The images presented to the models were however not blurred. Note that to make the subjects not identifiable we have blurred their images in Figures 1 and 4. The images presented to the models were however not blurred.

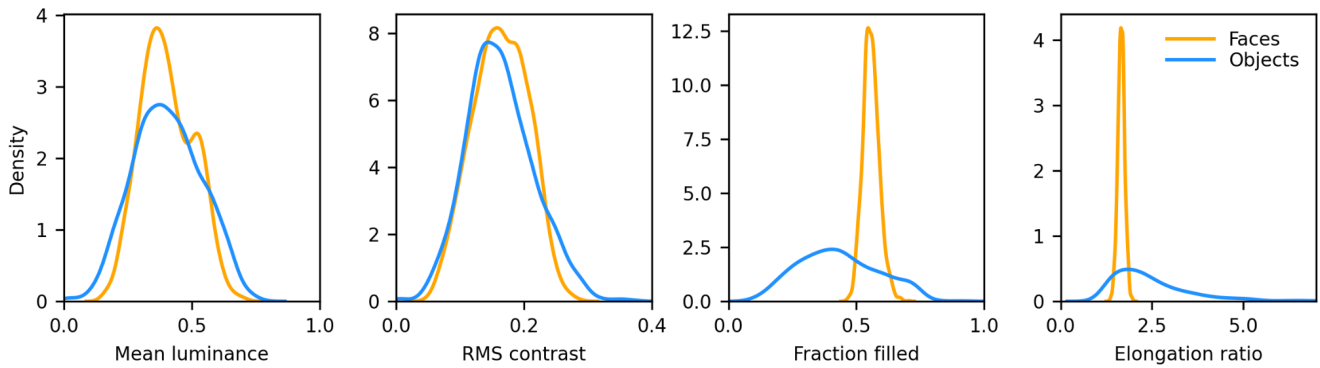

**Figure 2.** Distributions of low-level features for the faces and non-faces image sets. Mean luminance and RMS contrast of the images were well matched (Average mean luminance: face=102.96, non-face=103.07; Average RMS contrast: face=41.55, non-face=41.54). The fraction filled represents the fraction of pixels within the 256 x 256 area of the image proper (i.e., no background) that belong to the foreground (i.e., the face or non-face itself). Faces had a much narrower distribution, and a larger average (0.56 vs 0.43), than non-faces. The elongation ratio is the largest of the width/height or height/width ratios for the foreground. Faces had a much narrower distribution, and a smaller average (1.66 vs 2.51).

The images for the non-face dataset were selected by us from images freely available on the web for non-commercial use. We selected images of a wide range of animals with many different poses, fruits and vegetables, and man-made objects. We intentionally did not select any "unnatural" images, such as cartoons, flags, letters, and numbers, and we tried to add as many oval/circular shapes as we found. The images were of varying sizes and formats, and almost all were in color with a transparent background (we segmented those that did not have a transparent background to create one). We converted them all to grayscale and resized them (preserving their aspect ratio) so that their longest side was 246 px, and then centered them on a 256 x 256 px transparent canvas. We then computed the distribution of mean luminance and contrast of both sets of images. We found that their contrast levels overlapped, but non-face images were on average darker. We thus increased their luminance by 15% to approximately match the distributions. 8 samples from this set, centered on a 340 x 340 px mid gray background, are shown in Figure 1 (second row). The distribution of mean luminance, contrast, and two measures of shape (the fraction of foreground pixels within the original 256 x 256 px canvas, and the elongation ratio, defined as the ratio of the largest and smallest dimension of the rectangle that encompasses the foreground image) for the images we fed to the models are shown in Figure 2. Not surprisingly, the shape variation across the non-face set is much larger than across the face set.

Landscape images were selected from MIT's Places dataset of 10 million images (Zhou et al., 2018). We went through the dataset one image at a time until we had collected approximately 1,500 images that included neither humans nor animals, and depicted natural and urban scenes (for the latter mostly exteriors, with some interior scenes). Since all the images were in color, we then converted them to grayscale, resized them to 340 x 340 px, and corrected over and under exposed images by equalizing their luminance histogram using the OpenCV library (Bradski, 2000). 8 samples from this set of scenes are shown in Figure 1 (third row). A second set of landscape images was obtained from this set by reducing the contrast of each image by 50% and blurring it with a Gaussian filter with a standard deviation of 3 px. We refer to this set as blurred scenes. 8 samples from this set of blurred scenes are shown in Figure 1 (bottom row).

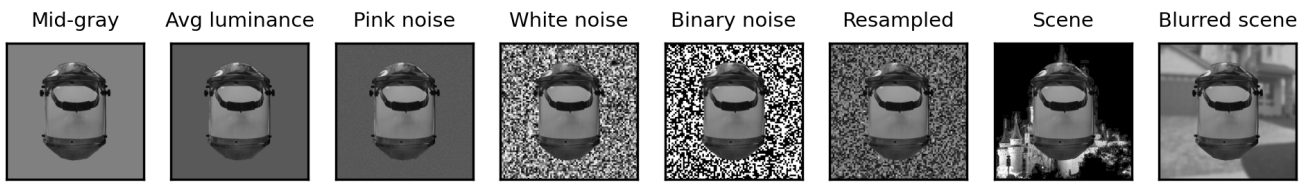

**Figure 3.** Superposition of a single non-face image on the backgrounds that we considered.

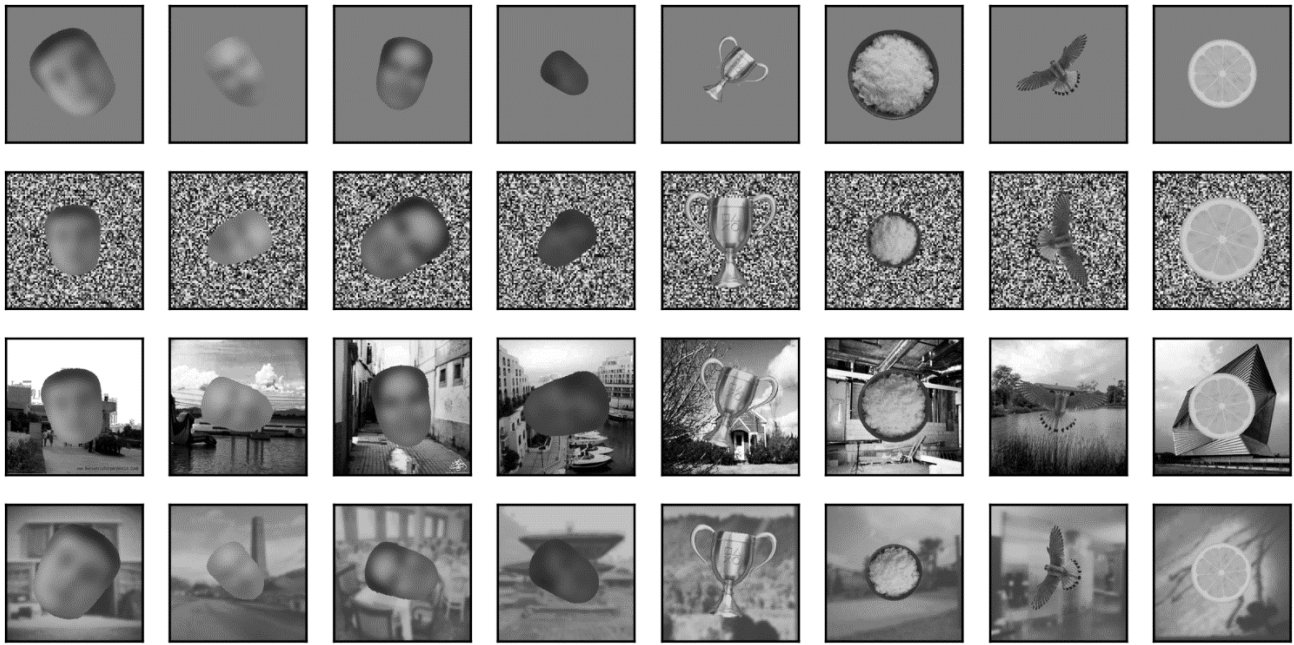

**Figure 4.** Samples of 4 faces and 4 non-faces on different backgrounds (from top to bottom: mid-gray, pixelated white noise, high contrast scene, blurred scene). Each image was randomly scaled down by up to 50% of its original size and rotated by up to 90° in either direction before being superimposed on the background. Here, but not in the simulations, face images have been blurred to prevent identification of the subject.

Images from the face and non-face sets were provided as inputs to the models (see below) one at a time, superimposed over a background canvas having a size of 340 x 340 px. Besides the high contrast and blurred scenes discussed above, the background could be a uniform mid gray (128), a uniform gray with the same luminance as the image presented, 2-D pink (1/f) noise, pixelated (4 x 4 px blocks) white noise (i.e., each block of pixels could have any luminance value from 0 to 255, randomly selected), pixelated binary noise (i.e., each block of pixels was set to 0 or 255, randomly), or pixelated noise with luminance values sampled from the image content. Samples of a single non-face image superimposed on these eight backgrounds are shown in Figure 3.

Besides being superimposed on the background, each image could be rescaled (always reduced in size, up to 50% in each dimension) or rotated (up to 180° in each direction). When in the text we refer to a scaling of images by x% we mean that each image was scaled by a random factor that varied between 100% and x%. Similarly, when we refer to a rotation of x° we mean that each image was rotated by a random angle between x° clockwise and x° counterclockwise. In Figure 4 we show 4 sample faces and 4 sample non-faces presented over four different types of background at a scaling of up to 50% and rotation of up to ±90°.

Finally, the location of each image on the background canvas was jittered (relative to the center) by a random number of pixels in the horizontal and vertical direction, up to ±42 px

In our last experiment, we tested our models on a gender classification task. For this task we manually classified the images from our face set into male or female, based on subjective appearance. This step was carried out on the original images, but the cropped versions of these images (same as for the face/non-face classification) were used as inputs to the models. A small fraction of the faces for which we felt unsure about the gender were excluded. We ended up with 924 images of males and 805 images of females, which we fed to our V1-like and AlexNet model (see below). For the FaceNet model we also created a set of images that were cropped to a section of each image that was identified by MTCNN (Zhang et al., 2016) as the bounding box for the face (the full horizontal extent of the face is usually kept, but the image is cropped vertically across the chin and the forehead). This region was then stretched (without preserving the aspect ratio) to match the image size required by FaceNet (160 x 160 px). In a final test we also used these images for AlexNet (resizing them to 224 x 224 px) or our V1-like model (resizing them to 256 x 256 px, centered on a 340 x 340 px mid gray background).

## V1-simple cells model

We implemented a two-layer model for binary classification in which the first layer was composed of a fixed set of oriented Gabor filters that resemble the receptive field properties of V1 simple cells (Marcelja, 1980; Jones and Palmer, 1987a). We used four sets of Gabor filters, each tuned to a different spatial frequency, spaced in one octave (i.e., a factor of two) steps. Since our images were scaled to fit in 256 x 256 px, we used for our largest scale 2-D Gabor functions with a carrier having a wavelength of 256 px. The wavelengths for the other three sets were then 128, 64, and 32 px. We set the standard deviation of the Gaussian envelope of the Gabor to 40% of the carrier wavelength, a value representative of orientation tuned cells in macaque V1 (Ringach, 2002) and cat area 17 (Jones and Palmer, 1987b). We truncated the size of each filter to 1.5 times the wavelength, again matching average properties of macaque V1 cells. At each scale we defined 16 different filters, covering 4 orientations (0°, i.e., vertical, 45°, 90°, and 135°) and four spatial phases (0°, 90°, 180°, and 270°) for the sinusoidal carrier. The filters for one scale are shown in Figure 5A (the top row shows the amplitude profile of the filters shown in the second row, those tuned to vertical orientation). Each set of 16 filters then tiled the entire space of the image (convolutional network), with a stride (distance in the horizontal or vertical direction between two nearby filters in image space) equal to half their wavelength. Because higher frequency filters have smaller sizes, they were more numerous than low spatial frequency filters. Overall, the model had 7,088 linear filters (distributed across the four spatial frequency channels as 144, 144, 1,024, and 5,776). Each filter was followed by a half-rectifying nonlinearity (negative values were set to zero, positive values were kept unchanged). The half-rectified outputs of the filters were then fed into a linear classifier, whose 7,088 trainable weights were initialized with the Kaiming uniform initialization method (He et al., 2015) (the bias term was initially set to zero). This process of summing the rectified outputs of the filters with learned weights is depicted in Figure 5B. Since we randomly jittered the location of the images by ±42 px, units with shorter wavelengths/higher spatial frequencies than those we used (e.g., 16 px) cannot be expected to extract any reliable information from the images, and thus would mostly contribute noise to the classifier.

## V1-complex cells model

We implemented a model of V1 complex cells by simply summing together the outputs of the V1 simple cells in the above model that shared the same spatial location, spatial frequency, and orientation (i.e., we summed the output of single cells that differed only in spatial phases, Figure 5C). This is a standard way of simulating V1 complex cells, analogous to the energy model of V1 complex cells (Carandini et al., 2005; Lian et al., 2021) but more biologically realistic (Hubel and Wiesel, 1962; Movshon et al., 1978; Martinez and Alonso, 2003). Because we considered four spatial phases at each scale and orientation, our V1-complex cells model had 1,772 units, whose outputs were fed to the linear classifier to learn the summing weights.

## V1 linear receptive field (RF) model

We also implemented a model of V1 simple cells linear receptive fields. Essentially, we took the model of V1-simple cells described above and removed the half-rectifying nonlinearity. Because with this modification the outputs of units with a spatial phase of 180° (270°) were simply the opposite of units with a spatial phase of 0° (90°), and thus entirely redundant (the weights of the classification layers can be positive or negative), we removed them. Thus, our V1 linear RF model had 3,544 units, whose outputs were fed to the linear classifier (Figure 5D).

## LGN-like model

We also implemented a two-layer model for binary classification in which the first layer was composed of a fixed set of difference of Gaussians (or Mexican-hat) filters that resemble the receptive field properties of neurons in the lateral geniculate nucleus (LGN) (Rodieck, 1965; Enroth-Cugell and Robson, 1966). Just as for the V1-like model, we used four sets of filters, each tuned to a different spatial frequency, spaced in one octave (i.e., a factor of two) steps. To match the preferred spatial frequency of the V1 filters, we set the standard deviation of the center Gaussian to 1/12th of the desired spatial frequency, the standard deviation of the surround Gaussian to 5 times that value, and the gain of the surround gaussian to 0.2 (the gain of the center Gaussian was 1). The size of the filters was the same as for the V1 filters. At each scale we defined 2 different filters, one mimicking on-center cells (i.e., cells that prefer a bright center and a dark surround) and the other mimicking off-center cells (i.e., cells that prefer a dark center and a bright surround). These filters are shown, for one spatial scale, in Figure 5E (the first row shows the amplitude profile of the filters shown in the second row). As with the V1 model, each set of 2 filters then tiled the entire space of the image

(convolutional network), with a stride (distance in the horizontal or vertical direction between two nearby filters in image space) equal to half their preferred wavelength. Because higher frequency filters have smaller sizes, they were more numerous than low spatial frequency filters. Overall, the model had 886 linear filters (distributed across the four spatial frequency channels as 18, 18, 128, and 722). Each filter was followed by a half-rectifying nonlinearity (negative values were set to zero, positive values were kept unchanged).The half-rectified outputs of the filters were then fed into a linear classifier (Figure 5F), whose weights were initialized with the Kaiming uniform initialization method (He et al., 2015) (the bias term was initially set to zero). As with the V1 model, using filters tuned to higher spatial frequencies would be unwarranted given our image jittering.

## AlexNet model

In some simulations we also used the model developed by Krizhevsky and colleagues for image classification (Krizhevsky et al., 2012). It is not the most sophisticated model available now, but it is the one that opened the ANN revolution in image recognition and it is still widely used as a benchmark. We used the model pretrained on the ImageNet classification task, which involves classifying images in one of 1,000 categories. We retained its stack of feature layers (which has 9,216 outputs), stripped it of its fully connected and classification layers, and replaced those layers with a single binary classification layer (identical to the one we used for our V1 model). Since AlexNet expects images of size 224 x 224 px, the 340 x 340 px images we used for our V1 model were rescaled to this size before feeding them to the model.

## FaceNet model

For the gender classification task we also used, as a benchmark, the model developed at Google by Schroff and colleagues for face recognition (Schroff et al., 2015). We used the Inception Resnet V1 model pretrained on 160 x 160 px images from the VGGFace2 face recognition task, which involves identifying a subject (out of a pool of 9,131 subjects, for which 3.31 million images are available). We retained its stack of feature layers (which in this model had 1,792 outputs), stripped it of its fully connected and classification layers, and replaced those layers with a single binary classification layer (identical to the one we used for our V1 model). Since FaceNet expects images of size 160 x 160 px and limited to the face bounding box extracted by MTCNN, the images we used for this model from our gender data set were pre-processed accordingly.

### Multi-tail AlexNet model

To quantify how the ability of AlexNet to extract useful features from our images varies as their processing progresses, we created a variant of AlexNet in which the classification is not based only on the final layer of features (as typically done and as we did above), but also on the output of intermediate layers. More precisely, we added a classification layer after the third, sixth, eighth, and tenth layers in the feature stack (in addition to the classifier that follows the 13th and last layer in the feature stack). Each binary classifier thus had a different number of inputs: classifier C1 had 46,656, C2 had 32,448, C3 had 64,896, C4 had 43,264, and C5 had 9,216 as before. These classification layers were trained concurrently but independently, and thus each maximized its ability to best predict the image based on the features it extracted.

### Model training and performance evaluation

For the face vs non-face task, we used the same procedure for all models. First, we randomly split each set of images (faces and non-faces) into two distinct sets, each with half the images (50/50 split). We then used one half for training the model, and the other half to evaluate its performance, which is thus fully cross-validated (the model was tested on images that were never seen by the model during the training phase). Unlike in classic model training tasks we did not feed the model identical images multiple times. Instead, we initially sampled, with replacement, 1,500 images from each of the faces and non-faces training subsets.

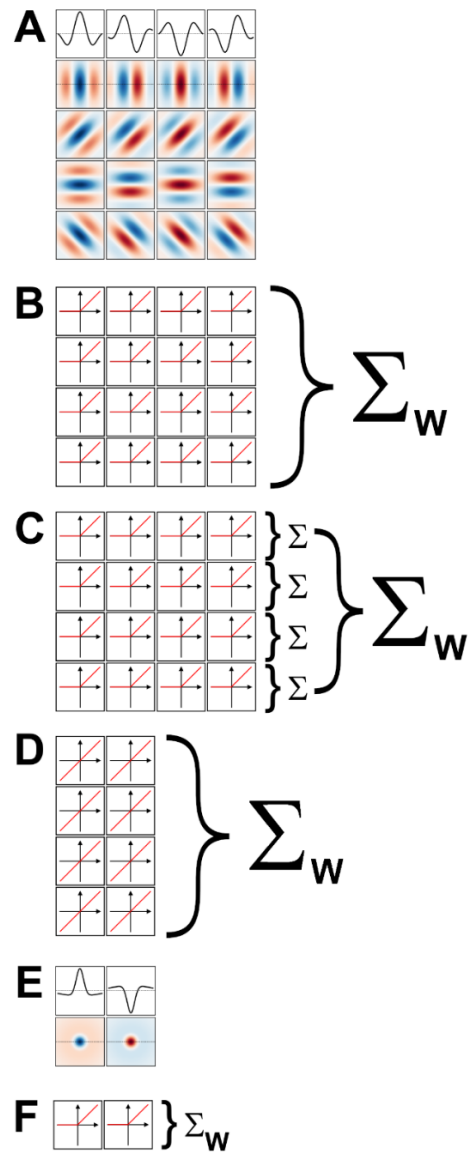

**Figure 5.** Models of early visual processing. A) Examples of oriented Gabor filters used for V1 model cells at one spatial location, and for one spatial scale. Four spatial phases (columns) and four orientations (rows) were used. Positive values are shown in shades of blue, and negative values in shades of red. The top row shows the amplitude profile of the filters in the second row. B) In a model of V1 simple cells the output of each filter is half-rectified, and then summed with a weight ($\Sigma$w) that is learned during model training. C) In a model of V1 complex cells the outputs of half-rectified filters associated with the same orientation, but different phases, are first summed ($\Sigma$w). The values for the various orientations are then summed with learned weights. D) In a linear model of V1 cells, only two phases are considered (the others are redundant), and their outputs are directly summed with learned weights. E) Difference of Gaussians filters used for LGN model cells at one spatial location, and for one spatial scale. On-center (left) and off-center (right) cells were used. F) In a model of LGN the output of each filter is half-rectified and then summed with a learned weight.

Before feeding each of these images to the model, they were jittered by a random amount and, when required, also rescaled and/or rotated by a random amount. Finally, a background was selected for the image (which, except for the mid gray background, would also be different for every image). Thus, if an image was present in the resampled training set twice, its location, scale, rotation, and background could/would be different in the two presentations. And in different epochs (training runs through the entire training sets) the images would thus also always be different. This can be seen as an extreme form of data augmentation, as the model was effectively never fed the same input twice. In practice this minimizes the risk of overfitting and maximizes model generalization. We empirically verified that performance on the testing set increased with the number of epochs, but quickly saturated, plateauing after 50 epochs, but improving only marginally after 10 epochs. Accordingly, we did not feel the need to use a validation set to prevent training from overfitting, and simply trained each model for 10 epochs.

After the model was trained, we sampled, with replacement, 1,500 images from each of the faces and non-faces testing subsets. We then fed these images through the model once and evaluated its performance (fraction of correctly classified images).

This entire process was repeated 100 times (100 different training/testing 50/50 splits), and we collected the performance for each round, thus obtaining a distribution of performance values. To summarize the performance of a model for a given condition (amount of scaling and rotation, and type of background) we report the median of the distribution of these 100 performance values. Another measure often used to quantify performance for binary categorization is d', which is computed as d' = Z(hit rate) – Z(false alarm rate), where Z is the inverse of the cumulative distribution of the standard normal distribution. Because in our study classification was on average symmetric (i.e., there were as many faces identified as non-faces as there were non-faces identified as faces) this can be simplified to d' = 2 x Z(hit rate).

For the gender classification task, since we had a considerably smaller image set, and some of the simulations (particularly those associated with the FaceNet model) were restricted to no or minimal data augmentation and thus a high risk of overfitting to the training set, we split the image set in non-overlapping training (50%), validation (10%), and testing (40%) sets. Training was then executed over up to 40 epochs and stopped based on the performance on the validation set (when performance failed to improve over 5 consecutive epochs). For the training and testing sets we sampled, with replacement, 400 images from each of the male and female training subsets. For the validation set we used 80 male and 80 female images (sampled without replacement) from the validation set. Since model performance was measured on the test set, whose images were never used during the training phase, it was fully cross-validated.

The models were implemented in Python 3.10, using the PyTorch 2.0 library. We trained the models to minimize the binary cross entropy with logits loss, the default choice for binary classification problems (Hopfield, 1987; Solla et al., 1988; Bishop, 1996). We trained the model using Stochastic Gradient Descent (SGD) with a learning rate of 0.01 and momentum of 0.9. Batch size was set to 64, and shuffling was applied during training.

# Results

## Face vs non-face classification

Our primary goal was to assess the impact of various factors (extent of scale variation, extent of orientation variation, and type of background) on the classification performance of our model of V1 simple cells. We started by fixing scale and rotation variation (at up to 50% and 90°, respectively) and evaluated the impact of 8 different types of backgrounds. Figure 4 shows samples of both faces (left four columns) and non-faces (right four columns) at these levels of scale and rotation variation, on four different backgrounds (each on a different row): mid-gray, pixelated white noise, high-contrast scenes, and blurred scenes.

For each type of background, we split the face and non-face image sets into two non-overlapping halves, and used one half (of each image set) for training the model, and the other half for testing its performance. The results are thus fully cross-validated, meaning that no images that were used to train the model were used to evaluate its performance. This entire process was repeated 100 times for each background type, so that we could assess the performance of the model over many different training/testing splits of the image sets. For each background we thus ended up with a distribution of performances.

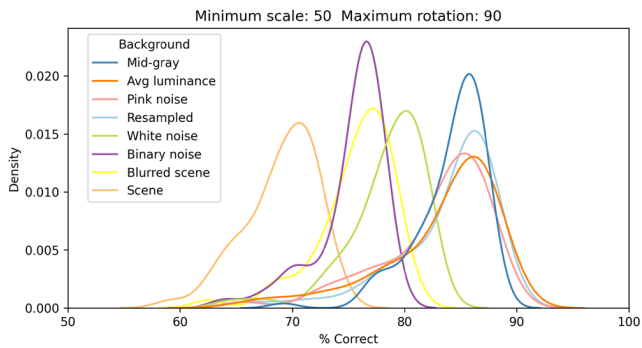

**Figure 6.** Distributions of classification performance of the V1 simple cells model as a function of the background used. In all cases each image was scaled down by up to 50% of its original size and rotated by up to 90° in each direction.

In Figure 6 we show the distributions of classification performance for the 100 training/testing sets used for each background. Gray backgrounds (either at a fixed mid value, Figure 4 top row, or at the average luminance of the image content), pixelated pink noise, and pixelated noise with luminance values matching the image content, were all associated with similar high performance (median performance 85%, 85%, 84%, and 85%, respectively). Using pixelated white noise as a background (Figure 4, second row) reduced performance considerably (79%). A blurred visual scene (Figure 4, bottom row) or pixelated binary noise (which had higher contrast than all background types we tested, Figure 3) made performance even worse (76% in both cases). Finally, high-contrast scenes (Figure 4, third row) had the most deleterious impact (69%).

To assess the impact of scale and rotation variability, we then selected four of these backgrounds (those shown in Figure 4), and for each of them varied independently scale across three levels (100%, i.e., no variation, 70%, and 50%) and rotation across five levels (0, i.e., no variation, 15°, 45°, 90°, and 180°). Performance degraded as variations in scale and rotation increased (Figure 7, top row), but the degradation was severe only for the largest variations and for the high contrast scene background. Because both faces and objects have a natural orientation, and are rarely tilted more than 45° from it, larger rotations are likely not representative of our daily experience. Similarly, pixelated and high-contrast backgrounds are not common, in part because backgrounds are often far from our plane of fixation and therefore defocused (i.e., have lower contrast) due to the limited depth-of-field of the human eye (Marcos et al., 1999). If we consider scale variations of up to 50%, and rotation variations of up to 45°, we see that with a gray background the median performance is 89%, and even with a blurred scene it is still a respectable 82%. While these levels of performance would not win any image classification contests, they are probably high enough to account for the preferential looking toward faces observed in human adults (Crouzet et al., 2010; Cerf et al., 2009) and infants (Fantz, 1961; Goren et al., 1975; Johnson et al., 1991), and in monkeys (Gothard et al., 2004; Sugita, 2008; Taubert et al., 2017).

These simulations show that a model with oriented filters like those observed in V1 simple cells is capable of extracting information that is useful, and potentially sufficient, to distinguish a face-like image from something that is unlikely to be a face (a coarse recognition task). We thus wondered if an even simpler model, with circular symmetric filters like those found in retinal ganglion cells and in the lateral geniculate nucleus (LGN), might achieve similar performance. However, when we trained and tested a model with LGN-like filters as we did for the V1-like model, we found (Figure 7, second row) that its performance was considerably worse, by as much as 20%. Thus, LGN-like filters seem to be much less useful for detecting face-like patterns, although they are sufficient for detecting areas of high contrast.

Next, we wondered how much of the performance of the V1 model could be accounted for by its sensitivity to the spatial phase of its filters, a characteristic of V1 simple cells. We thus implemented a model of V1 complex cells, which are insensitive to spatial phase. We found (Figure 7, third row) that its performance closely matched that of the model of V1 simple cells, underperforming by a small margin across the set of conditions tested, indicating that spatial phase plays only a minor role.

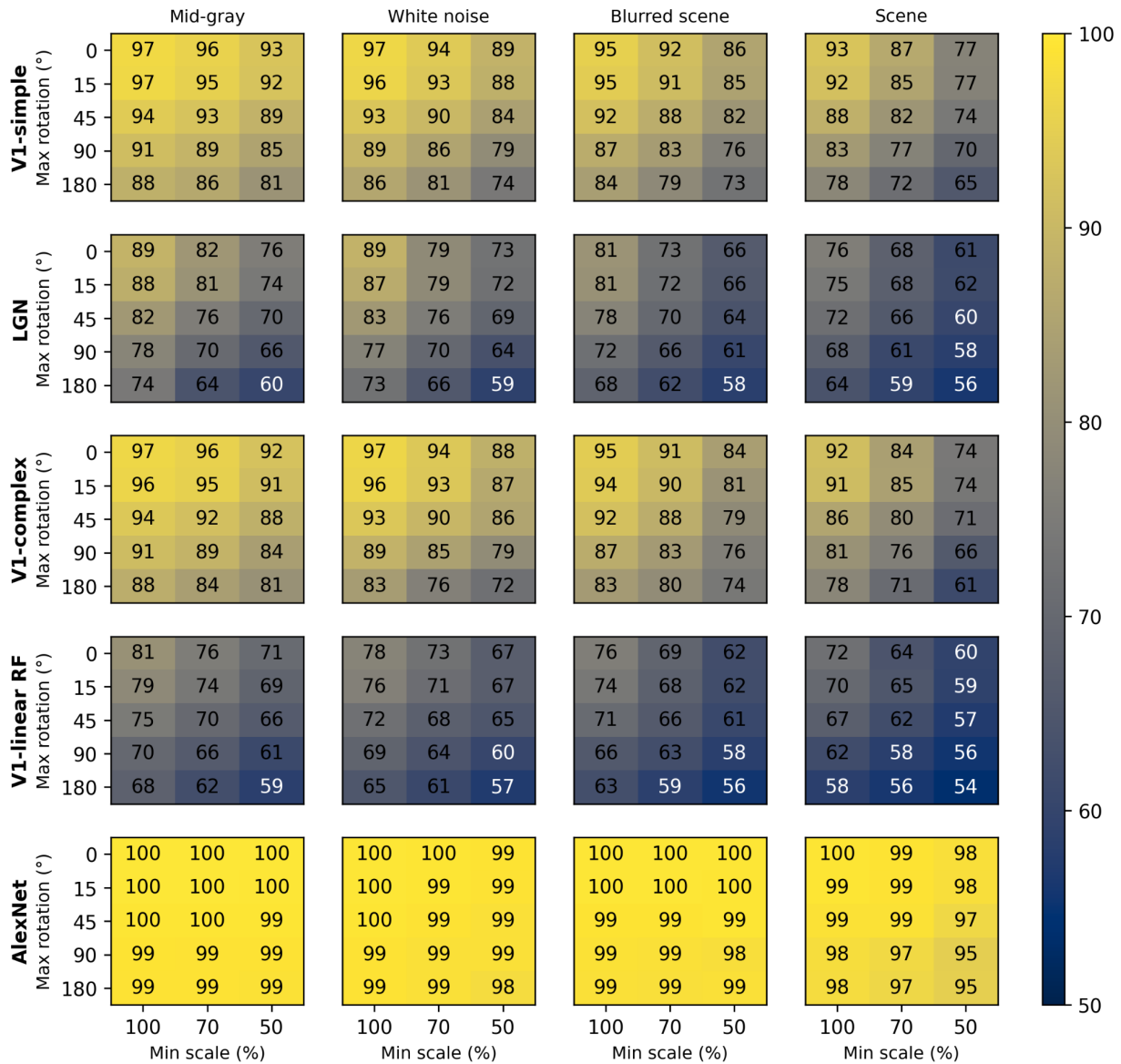

**Figure 7.** Median classification performance of our models (from top to bottom: V1 simple cells, LGN, V1 complex cells, V1 linear RF, and AlexNet) as a function of the background (each panel), scale (three levels across columns), and rotation (five levels across rows).

Finally, to estimate the contribution of the rectifying nonlinearity to classification performance, we evaluated a model of V1 simple cells linear receptive fields. This is essentially our model of V1 simple cells, but with only two spatial phases represented at each location (because the other two become redundant) and without the rectifying nonlinearity. We found that this model underperformed the V1 simple and complex cell models by large margins (Figure 7, fourth row), and was even worse than the LGN model. This clearly highlights the critical importance of nonlinearities, even in models as simple as those presented here, and counters our tendency to intuitively attribute the lion's share of effects to the linear filter.

Some might be surprised by how well these models performed. To provide an additional reference, and a reality check, we also tested a model that has been compared in the past to the primate ventral visual pathway (Cadieu et al., 2014). This model was developed by Krizhevsky and colleagues and won the ImageNet image classification contest in 2012 (Krizhevsky et al., 2012), and is usually referred to as AlexNet (after Krizhevsky's first name). We selected a version of this model that was already trained on the ImageNet classification task, in which an image must be categorized as belonging to one of 1000 categories. We then froze the weights of its feature maps (which can be loosely seen as a cascade of several of the filter layers that we used in our V1- and LGN-like models, but with many more, and more complex, filters), and replaced its classification layers with a linear classifier, identical to the one we used for the V1- and LGN-like models (although with a different number of inputs, corresponding to the number of filters in the last feature layer of AlexNet). Not surprisingly, we found that AlexNet does a much better job compared to our simple V1-like model, and it is much less sensitive to changes in scale, orientation, or background (Figure 7, bottom row). Even with our most challenging condition (up to 50% scaling, up to 180° rotation, high-contrast scene background) its median cross-validated performance is 95% (compared to 65% for the V1-like model); with up to 50% scaling, up to 45° rotation, and a blurred scene background it performed at 99% (compared to 82% for the V1-like model).

To allow for a more direct comparison between the performance of the various models, in Figure 8 we directly compare the performance of the V1-simple cells model to the that of the other models. Instead of using the percent correct measure that we have use throughout the paper, here we used the sensitivity index d' (see Methods), another commonly used metric to characterize binary discrimination (which we clipped at 5.15, corresponding to 99.5% correct). Easy (hard) tasks are associated with high (low) values of d'. Each data point is associated with one condition from Figure 7 (3 scaling factors, 5 rotation factors, 4 backgrounds, thus 60 points for each model), and we plot on the x-axis the value of d' observed with the V1-simple cells model and on the y-axis the value of d' observed with one of the other models (see legend) for the same condition. Points below

(above) the diagonal indicate that the V1-simple cells models performed better (worse) than the other model. Disregarding values at or near saturation at high and low values of d', when expressed in units of d' the relative performance of the various model is relatively constant as difficulty varies (i.e., the points for each model are distributed along lines parallel to the main diagonal).

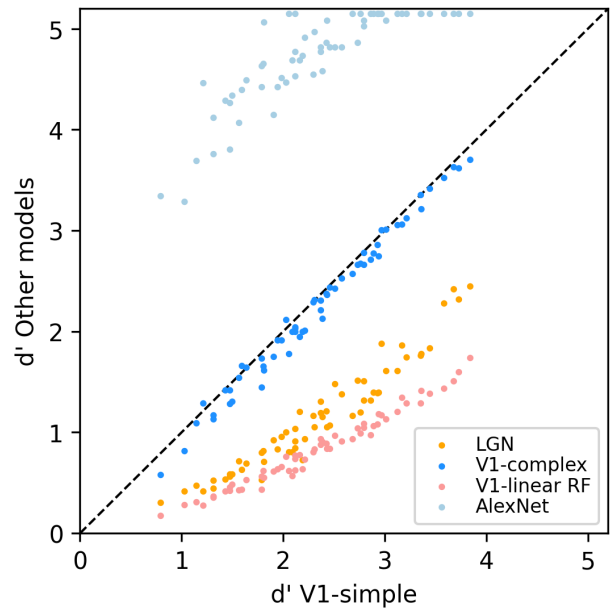

**Figure 8.** Scatter-plot of the median sensitivity index d' of our models. The V1 simple cells model is used as a reference and is plotted on the x-axis, and all other models we tested (see legend) are plotted against it (y-axis).

## Probing the V1-like model through selective lesions

So far, we have shown that a V1-like model (whether based on responses of simple or complex cells) contains more (linearly separable) information about distinguishing a face from a non-face object than a LGN-like model, but much less than a modern ANN could extract from the same images. We also showed that the linear receptive field and the output nonlinearity are both critical determinants of performance. We next sought to gain further understanding about which units of the V1-like models are most useful to the classifier. In a manner that is akin to what is done in neuropsychology, we thus simulated "lesions" of the model by removing subsets of model units, and evaluated their impact on the ability of the model to classify our stimuli. Since our V1-like models have four spatial frequency channels, we first eliminated from our model of V1 simple cells one or more of them and refit it to the data, for the same amount of scaling (up to 50%), rotation (up to 45°), and type of background (blurred scene). We found (Figure 9, left panel) that, compared to the full model (blue), dropping any single channel (yellow) had a minimal impact on performance. Dropping two channels (purple) also had a small impact, unless the two channels tuned to the higher frequencies were both dropped. Finally, when a single channel was kept (pink), we found that the two highest channels by themselves can perform essentially as well as the single model, whereas the two tuned to lower SFs are significantly impaired. The lowest spatial frequencies, which might have been thought to be associated with the overall shape of the face, thus do not play a major role in determining performance under these conditions (although there is the potential confound that the number of filters increases with spatial frequencies, see Methods). When images are presented with no random changes in scale and orientation, and on a gray background, the two lowest SF channels can effectively classify our dataset (Figure 9, right panel), indicating that their usefulness decreases as image variation increases.

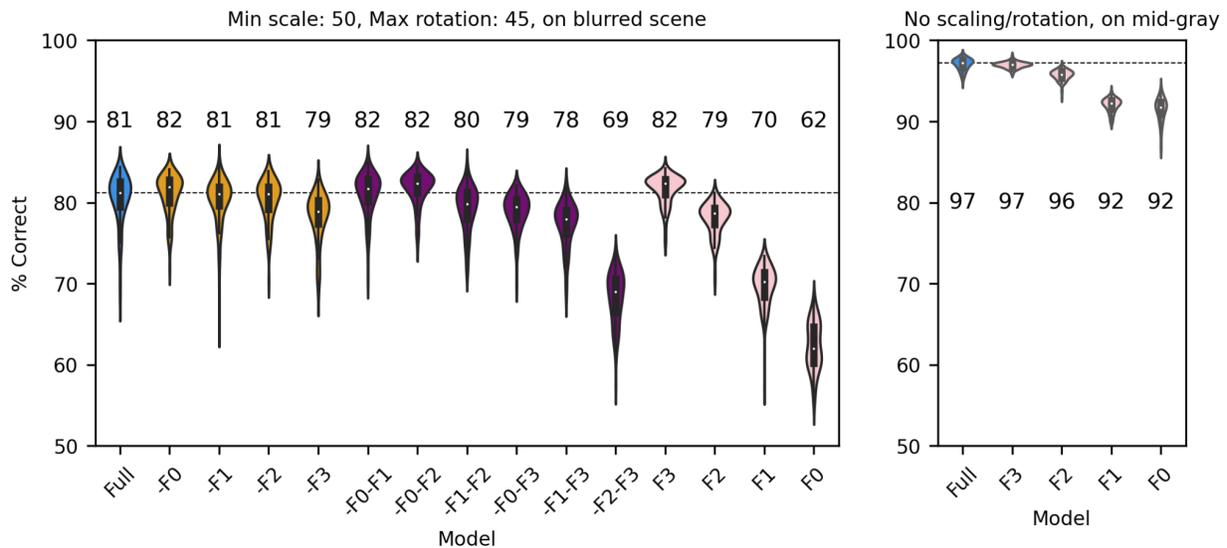

**Figure 9.** Classification performance of the full model of V1 simple cells (blue) and versions of the model in which one (yellow), two (purple) or three (pink) SF channels have been removed. Median performance is shown for each distribution. Left panel: Each image was scaled down by up to 50% of its original size, and rotated by up to 45° in each direction, and was presented on a blurred scene background. Right panel: We presented the original images (no scaling or rotations) on a mid-gray background to the full model or versions of it with a single spatial frequency channel.

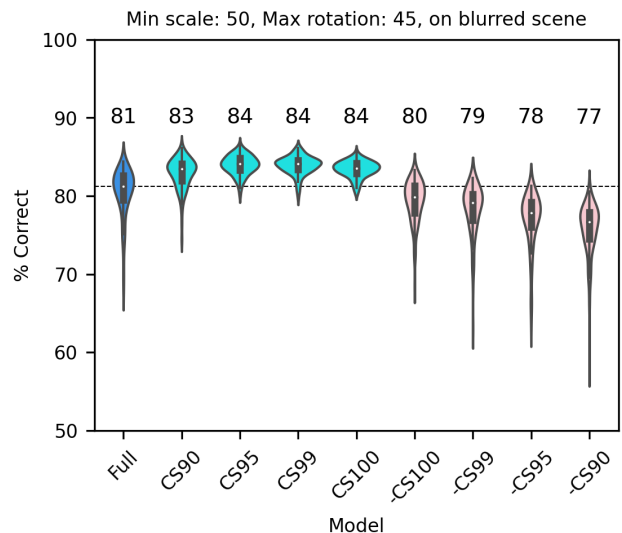

**Figure 10.** Classification performance of the full model of V1 simple cells (blue) and versions of the model in which we kept only those filters that were either consistently associated with positive or negative weights over a fraction of runs of the full model (cyan, at least 90, 95, 99 or 100 runs), or in which those same filters were omitted (pink, note that the order is reversed) In all cases each image was scaled down by up to 50% of its original size and rotated by up to 45° in each direction, and was presented on a blurred scene background. Median performance is shown above each distribution.

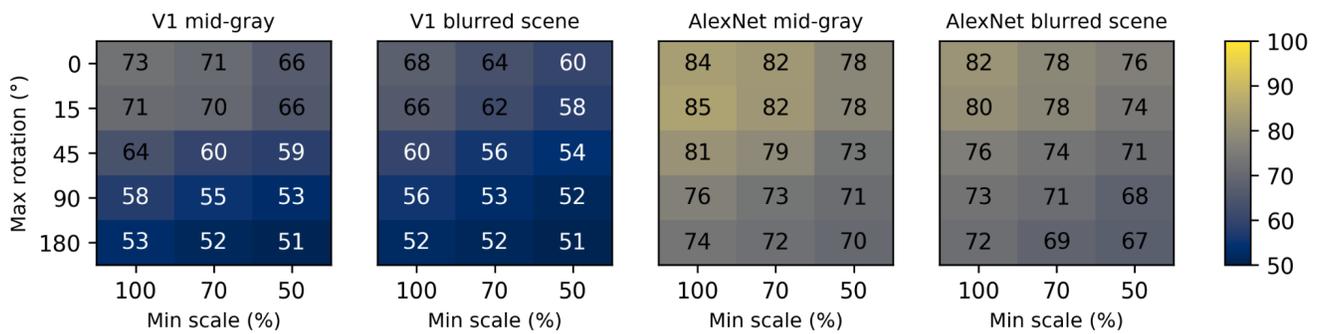

**Figure 11.** Median classification performance of the V1-like (simple-cells) and AlexNet models in the gender classification task. Two different backgrounds were tested for each model, varying scale (three levels across the x axis) and rotation (five levels across the y axis) as before.

Next, we considered a set of lesions that targeted the most consistent inputs to the classifier. We kept or removed only those filters that in the trained full model of V1 simple cells were associated with a classifier weight that was positive or negative on at least: 90/100 model runs (CS90, 1686 filters); 95/100 model runs (CS95, 1162 filters); 99/100 model runs (CS99, 612 filters); or all 100 model runs (CS100, 406 filters). Note that filters were kept (or removed) across all SF channels (e.g., for CS90, 22% of filters in F0, 38% of filters in F1, 32% of filters in F2, and 20% of filters in F3 were kept; for CS100 the percentages were 2%, 9%, 10%, 5%, respectively) although in absolute numbers many more filters were kept (or removed) from the higher SF channels. In all cases in which we kept only the most consistent weights (Figure 10, cyan), the model performed as well as the full model (or slightly better than it, probably because filters that contributed mostly noise were omitted). What was more surprising, removing the filters associated with these weights (Figure 10, pink) resulted in only a relatively small drop in performance, indicating that there is a fair amount of redundancy in the model, which allowed the remaining filters to compensate for the loss. However, an inspection of the scatter of model performance across runs (which can be visually estimated by the length of each violin in Figure 10) reveals that keeping only the most relevant weights has a strong impact on the reliability of the classification, presumably by making the classifier easier to train (i.e., less prone to settle on a local minimum).

Applying these same analyses to our model of V1 complex cells yielded similar results (not shown).

## A fine recognition task: Gender classification

While it is difficult to understand what exactly the V1-like model responded to (a possibly ill-posed question), the large within-group shape similarity across faces is an obvious candidate (Rosch et al., 1976). To evaluate the V1-like model under more challenging conditions, requiring finer discrimination abilities, we tested it on a gender classification task. We subdivided our face set into two groups (male and female), and then trained the V1-like and the AlexNet models as we did before for the face vs non-face classification (see Methods). We only tested two backgrounds, mid-gray and blurred scenes. Not surprisingly, cross-validated performance was considerably worse than before (Figure 11), dropping by approximately 25% points for the V1-like model and by 20% point for AlexNet. The performance of the V1-like model stayed above 70% only for small variations in scale and rotation; at the reference condition we considered before (up to 50% scaling, up to 45° rotation, blurred scene background) the performance was essentially at chance level, compared to the 82% that we observed in the face vs non-face task.

We also tested FaceNet (Schroff et al., 2015) in this task, a model developed explicitly for face recognition, as opposed to image classification. We expected it to perform much better than the other two models in this task, and indeed its cross-validated classification performance on our data set was 93%. However, there is a catch: This network is not trained to work directly on the images, but instead it first processes the images with another ANN, called MTCNN, which extracts a bounding box containing a face, and then this part of the image is stretched to the size preferred by FaceNet (160 by 160 px) and fed to it. There is thus a single scale, no rotation, and very little background. A fairer comparison would thus involve feeding to our two other models the same images that we fed to FaceNet. When we did that the V1-like model performed at 91% correct, and AlexNet matched FaceNet at 93% correct. Note that these performances are much better than those we obtained with no scaling or rotation on our image set on a mid-gray background (73% and 84%, respectively). What accounts for this difference? One possibility is the stretching of the images operated by MTCNN. Another possibility is the lack of jitter in the images - recall that in all previous tasks the images were always randomly jittered by up to 42 px in all directions. To find the answer we again classified the original images with the V1-like and AlexNet models, this time omitting jittering, for the no scaling, no rotation, and mid-gray background condition. The cross-validated performance of both models was now 90%, indicating that the lack of jitter accounts for the vast majority of the difference observed. Obviously the presence of jitter limits, for both our V1-like model and for AlexNet, the ability to perform fine (but not coarse) discriminations.

## Object recognition along the visual hierarchy

We have shown here how recognition performance differs, for two tasks, in models of the earliest stages of visual processing (LGN, V1 simple and complex cells) and in a widely used ANN, whose final stage has been likened to IT processing. It would of course be interesting to understand how performance varies along the entire ventral visual pathway. Unfortunately, there is not yet consensus on how the properties of V2, V3, V4, and IT emerge from their inputs. However, since it has been argued that there is a hierarchical similarity, in terms of feature selectivity of individual units, between layers of ANNs and areas of the primate visual system, we evaluated how classifiers trained on the features extracted by intermediate layers of

AlexNet fared in our tasks (Figure 12A). While direct comparisons have not been carried out, in our multi-tail version of AlexNet one can roughly consider the stack of layers F1 as akin to area V1, F2 to V2, F3 to V3, F4 to V4, and F5 to IT. In Figure 12B we plot, for a model trained on a face vs non-face classification task (with images presented on a blurred background, scaled by up to 50% and rotated by up to 45°), the distribution of classifier outputs when presented with non-faces (blue) or faces (orange) from the testing image set. While performance improved as we moved deeper in the feature stack, classifier C1 already provided actionable information, and going from C2 to C5 provided a very limited improvement in performance for the additional computational load (for computers) and time (for the brain and computers) required. In Figure 12C we plot the results for a model trained on a gender discrimination task (with images presented on a blurred background, scaled by up to 50% and rotated by up to 45°). This is a much harder task, and shortcuts incur significant costs, making waiting for the output of C5 a sensible strategy. For all their limitations, these simulations again highlight the ability of earlier stages of processing to perform adequately in tasks requiring coarse, but not fine, categorization.

## Discussion

The last ten years have seen a resurgence in the study of object recognition, both in computer science and in primate neuroscience. What often goes unacknowledged is that almost all the focus is on fine discrimination, or what in primates could be termed foveal object recognition. We have shown here that for simpler discrimination tasks signals in early visual areas are sufficient to achieve reasonable accuracy. In itself this is not a new finding (Serre et al., 2007), but it is one that has been sidelined once the limitations of early visual areas to support fine object recognition under large variations in orientation, illumination, and background (such as recognizing the image of an elephant upside-down over a high contrast skyline of New York) were revealed ((Pinto et al., 2008) (Rajalingham et al., 2018)).

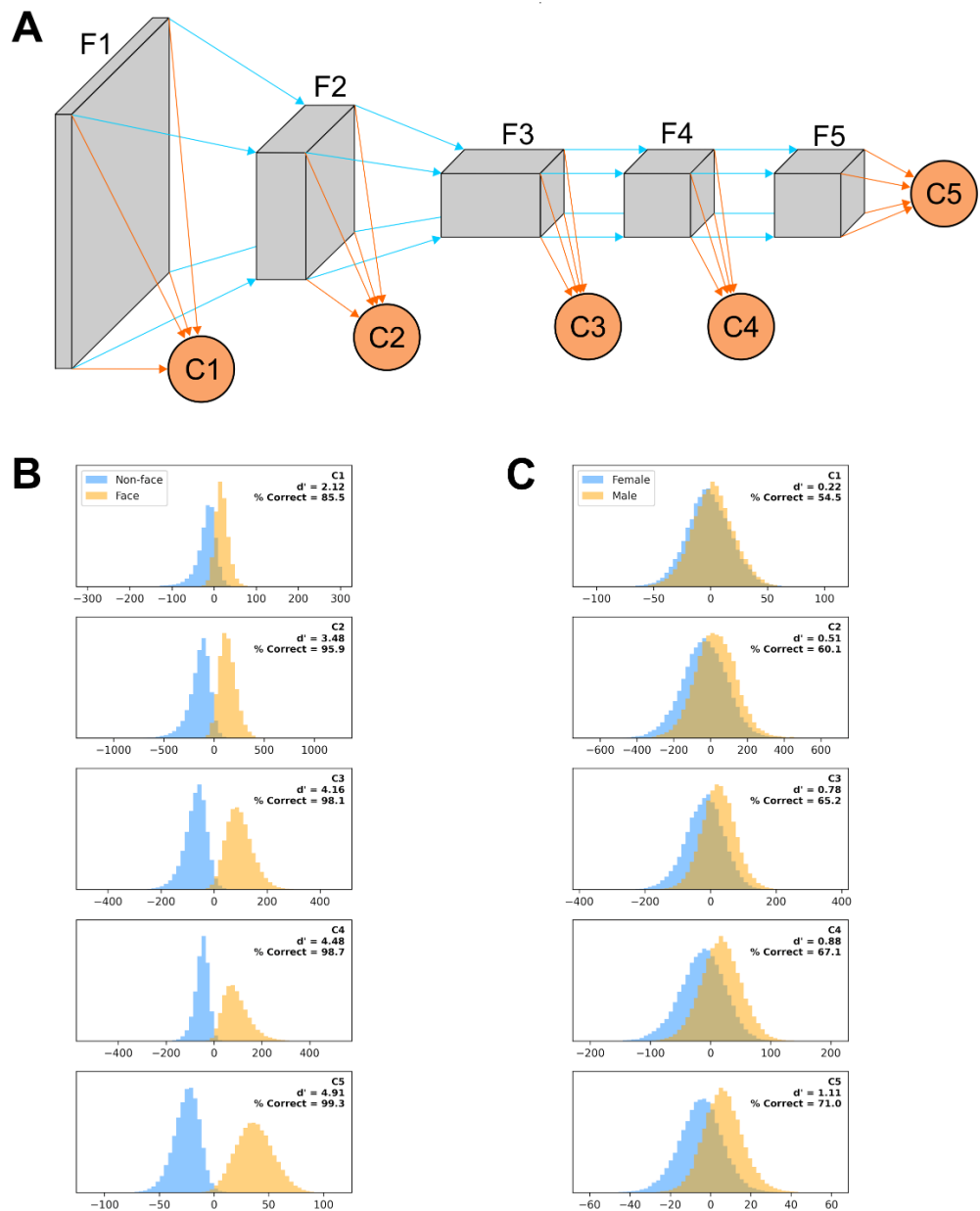

**Figure 12.** A: A "multi-tail" (our nomenclature) version of AlexNet, in which the feature stack has been divided into five stages, and a binary classification layer is appended to each stage (see Methods). B: Distribution of each classifier outputs when presented with non-faces (blue) or faces (orange) from the testing image set for a model trained on a face vs non-face classification task with images presented on a blurred background, scaled by up to 50% and rotated by up to 45°. The outputs are collated across 100 different training/testing splits of the data. C: Same as panel B, but for a model trained on a gender discrimination task (with images presented on a blurred background, scaled by up to 50% and rotated by up to 45°).

Our main contribution here is to more accurately describe these limitations for a task that is behaviorally important and that had not been considered in previous studies. Furthermore, by comparing the performance of different variations of our models we were able to highlight the relative contributions of their various parts.

Unlike computer science models, the brain of primates must contend with visual inputs that do not uniformly sample the visual field, making object recognition in the periphery an altogether different problem from object recognition in the fovea. There is no doubt that foveal vision is critical to primates' ability to survive and thrive, as clearly demonstrated by the fact that up to 30% of area V1 is devoted to this region (Perry and Cowey, 1985; Azzopardi and Cowey, 1996), and by the devastating effects of macular degeneration (Fleckenstein et al., 2021). However, some fundamental aspects of object recognition are overlooked when the focus is on foveally presented objects in an unusual pose against an out-of-context high contrast background (what has been termed core object recognition, (DiCarlo and Cox, 2007; DiCarlo et al., 2012)). Most of our daily experience is quite different, and involves quickly localizing familiar objects in familiar environments, across the visual field (Eckstein, 2017). Because the fovea covers less than one part in 5000 of our visual field, without a fairly sophisticated system to detect objects in the periphery, our ability to make good use of IT's capacity for fine object discrimination would be severely hindered.

Recognizing that an object of interest is somewhere in the periphery is not sufficient: its precise localization is critical. Hence, peripheral recognition should be based on neurons that carry position information. It has been argued that, in tasks involving difficult discriminations of objects on high contrast backgrounds, the available position information increases along the visual stream (Hong et al., 2016), and is thus higher in IT than in V1. However, that study used a rate code to extract such information (where a larger value of a single output indicates a more eccentric position); such a code would be difficult to use for controlling behavior, as areas of the salience network (such as LIP, FEF, and SC) are retinotopically organized and interconnected. In contrast, a retinotopic code (where different cells are active to indicate the presence of an object at different retinal locations) would be much more readily usable, but this information is progressively diluted deeper (i.e., more anteriorly) in the hierarchy of visual areas. Object recognition based on signals in retinotopically organized early visual areas is thus not only possible (as we have shown here) but would also be eminently useful in primates. Importantly, early visual areas do project in a retinotopically organized manner to areas of the salience networks, such as the SC (Cerkevich et al., 2014). In afoveated animals, which devote to vision a smaller fraction of their brain, this kind of rudimentary object recognition might very well comprise the entirety of visual object recognition, as they might rely more on motion signals and other sensory modalities to direct their behavior (motion is of course an important visual cue also for primates, but it normally plays a secondary role relative to form vision, with some important exceptions, e.g., breaking camouflage, biological motion).

Here we focused on face recognition, because it is one of the most widely studied visual classification tasks, and it is one at which primates excel (Diamond and Carey, 1986; Carey, 1992; Pascalis and Bachevalier, 1998; Pascalis et al., 1999; Rosenfeld and Van Hoesen, 1979). It is also notable that a preference for looking at faces is already present in newborns (Pascalis and de Schonen, 1994; Goren et al., 1975), well before the development of the cortical machinery that supports face identification in older children and adults, the so-called face-patch network (Livingstone et al., 2017). This observation had led to the suggestion that preferentially looking toward faces might initially be mediated by direct projections from the retina to the SC, with cortical projections to the SC taking over during development (Morton and Johnson, 1991). Indeed, SC cells have been found to respond more strongly to images of faces than non-faces in adult monkeys (Nguyen et al., 2014; Yu et al., 2023). Our finding that circular center-surround filters like those found in retinal ganglion cells and LGN are not well-suited for this task argues against a retinal source for this information. However, because it is difficult to calibrate the classification performance of our models against the preferential looking methods used when studying infants, it remains possible that even the limited performance of our LGN model might be able to account for the preferences of infants.

We tested our models also on gender classification. We chose this task because it requires discrimination of fine features, and yet it has been shown that it requires less time than face identification (Dobs et al., 2019), and it is preserved in prosopagnosic patients (Sergent et al., 1992), whose cortical face processing network (Tsao et al., 2003; Kanwisher and Yovel, 2006) is known to be disrupted (Rosenthal et al., 2017). These findings point to gender classification as possibly relying on visual cortical signals that precede the face-network in IT. Our results indicate that V1 is an unlikely source of such signals. However, since both texture and curvature have been proposed to play important roles in gender discrimination (Brown and Perrett, 1993; Bruce et al., 1993; Hole and Bourne, 2010), area V2, which is specialized for processing naturalistic textures (Freeman et al., 2013; Movshon and Simoncelli, 2014), and area V4, which is specialized for processing curvature (Pasupathy and Connor, 2002; Nandy et al., 2013), might carry useful signals. V2 and V4, like V1, project directly to the SC (May, 2006), and would thus also be well-positioned to bias refixations.

## A different architecture for object recognition

The ideas about object classification and recognition that dominate the field, with their emphasis on foveal inputs and context-independent processing, are certainly important, but they omit some crucial functional aspects of how our specialized primate visual system identifies objects. A unitary hierarchical architecture of ever more complex feature detectors with a read-out/classification stage at its end, similar in structure to ANNs for object recognition (with only C5 in Figure 12), might work well in computer science, but it is a brittle architecture, at odds with evolutionary principles and experimental evidence. We envision a parallel architecture, in which classification is based on signals in multiple visual areas (as in Figure 12), but, unlike in ANNs, with learning for coarse recognition separately at each retinal location. Such a system would be more flexible, has strong redundancy, and it is easier to see how it could have evolved from simpler systems. Information from some stages might be available earlier than that from others, but might also be less reliable, yielding a speed/accuracy trade-off that could be easily arbitrated, as we saw in our multi-tail version of AlexNet (Figure 12). The information coming from each area could be weighted differently in the center and in the periphery, and also by pathways involved in action (e.g., controlling refixation through the saliency network) vs perception. Such a system would be more flexible than a classic ANN-like architecture, in which the loss of any part/stage would be devastating, requiring retraining of the entire model (as in our simulated lesions). It would be akin to the "wisdom of the crowd" (Aristotle, 2013; Saha Roy et al., 2021; Madirolas et al., 2022) or "ensemble learning" (Hansen and Salamon, 1990; Schapire, 1990; Polikar, 2006) architectures, in which a highly complex decision process is replaced by many simpler ones, each evaluating the evidence independently and reaching its own conclusion. If any of the processes becomes unavailable, the others carry on, as is observed in real life (Matsumoto et al., 2016).

## Acknowledgements

This work was supported by the Intramural Research Program of the National Eye Institute. We thank Prof. Doris Tsao for kindly sharing with us a carefully curated diverse set of frontal images of human faces. We thank Drs. Leor Katz, Gongchen Yu, and Corey Ziemba for many insightful discussions and for comments on a previous version of the manuscript.